\documentclass[aps,twocolumn]{revtex4}
\usepackage{graphicx}

\usepackage{graphicx}

\newcommand{\be}{\begin{equation}}
\newcommand{\ee}{\end{equation}}
\newcommand{\bqn}{\begin{eqnarray}}
\newcommand{\eqn}{\end{eqnarray}}

\begin{document}

\title{Thermal Effects on the Stability of Excited Atoms in Cavities}

\author{F. C. Khanna}
\email[E-mail address: ]{khanna@phys.ualberta.ca}
\affiliation{Theoretical Physics Institute, Department of Physics,
University of Alberta, Edmonton, AB T6G 2J1 and TRIUMF, Vancouver,
BC V6T 2A3, Canada}
\author{A. P. C. Malbouisson}
\email[E-mail address: ]{adolfo@cbpf.br} \affiliation{Centro
Brasileiro de Pesquisas F\'{\i}sicas/MCT, 22290-180, Rio de
Janeiro, RJ, Brazil}
\author{J. M. C. Malbouisson}
\email[E-mail address: ]{jmalboui@ufba.br} \affiliation{Instituto
de F\'{\i}sica, Universidade Federal da Bahia, 40.210-310,
Salvador (BA), Brazil}
\author{A. E. Santana}
\email[E-mail address: ]{asantana@fis.unb.br}
\affiliation{Instituto de F{\'\i}sica, Universidade de
Bras{\'\i}lia, 70910-900, Bras{\'\i}lia-DF, Brasil}

\begin{abstract}
An atom, coupled linearly to an environment, is considered in a
harmonic approximation in thermal equilibrium inside a cavity. The
environment is modeled by an infinite set of harmonic oscillators.
We employ the notion of dressed states to investigate the time
evolution of the atom initially in the first excited level. In a
very large cavity (free space) for a long elapsed time, the  atom
decays and the value of its occupation number is the physically
expected one at a given temperature. For a small cavity the excited
atom never completely decays and the stability rate depends on
temperature.


\end{abstract}


\maketitle

\section{Introduction}

Inhibition of spontaneous emission by confined atoms is a well-known
phenomenon, currently related to the dipole orientation with respect
to parallel mirrors or to the relation between the size of the
confining device and the emission wavelength (see for
instance~\cite{farina1,farina2} and other references therein). The
theoretical understanding of these and other effects in atomic
physics on perturbative grounds requires the calculation of very
high-order terms in perturbative series, that makes the standard
Feynman diagram technique practically unreliable. This has lead to
trials of treating non-perturbativelly such kind of systems using
the semi-quantitative idea of a dressed atom~\cite{bouquinCohen}.
However serious difficulties, due to  nonlinearity, are present to
get rigorous results in these approaches. A way to circumvent these
mathematical difficulties is to assume that under certain conditions
the coupled atom-electromagnetic field system may be approximated by
a system composed of a harmonic oscillator coupled  linearly to the
field modes through some effective coupling constant $g$. This is
the case for linear response theory in $QED$, where the electric
dipole interaction gives the leading contribution to the radiation
process~\cite{Wylie,Jhe}. Although a linear model is a simple
theory, it permits a better understanding of the need for
non-perturbative analytical treatment of coupled systems. This is
the basic problem underlying the idea of \textit{dressed} quantum
mechanical operators.

The perturbative  treatment of interacting systems is carried out by
considering bare, non-interacting fields. The interaction is taken
into account order by order in powers of the coupling constant.
However there are situations where the use of perturbation theory is
not reliable, as in the low energy domain of quantum chromodynamics
and resonant effects in atomic physics, associated with the coupling
of atoms with strong radiofrequency fields~\cite{bouquinCohen}.

The idea of a bare particle associated to a bare matter field is
actually an artifact of perturbation theory. A charged physical
particle is always coupled to the gauge field, i.e, it is always
``dressed" by a cloud of quanta of the gauge field (photons, in the
case of electrodynamics). In a simplified model for a radiating
atom, a way to treat directly dressed objects, has been introduced.
This is the method of \emph{dressed} states and {\it dressed}
coordinates~\cite{adolfo1} that has been employed in several
cases~\cite{adolfo2,adolfo3,adolfo4,adolfo5,adolfo6}.

In this paper we generalize the zero-temperature formulation,
dealing non-perturbatively with the inhibition of spontaneous
emission~\cite{adolfo2,adolfo3}, to finite temperature. The
objective is to consider  the stability of confined excited atoms in
an environment at finite temperature.

\section{The model}

We start by considering a bare atom approximated by a harmonic
oscillator described by the bare coordinate and momentum $q_0, p_0$
respectively, having \emph{bare} frequency $\omega _0$, linearly
coupled to a set of $N$ other harmonic oscillators (the environment)
described by bare coordinate and momenta $q_k, p_k$ respectively,
with frequencies $\omega _k$, $k=1,2,\ldots ,N$. The limit
$N\rightarrow \infty$ will be  taken later. A model of this type,
describing a linear coupling of a particle with an environment, has
been used for years in several situations, for instance  to study the quantum Brownian
motion of a particle with the path-integral
formalism~\cite{zurek,paz,caldeira,caldeira1}. The whole system is
supposed to reside inside a spherical cavity of radius $R$ in
thermal equilibrium with the environment, at a temperature
$T=\beta^{-1}$ ($k_B$, the Boltzmann constant is taken equal to 1).

The Hamiltonian for such a system is written in the form,
\begin{equation}
H=\frac 12\left[ p_0^2+\omega _0^2q_0^2+\sum_{k=1}^N\left(
p_k^2+\omega _k^2q_k^2\right) \right] - q_0\sum_{k=1}^Nc_kq_k,
\label{Hamiltoniana}
\end{equation}
where the $c_{k}$'s are coupling constants.
In the limit $N\rightarrow
\infty $, we recover the case of an atom coupled to the
environment, after redefining divergent quantities, in a manner
analogous to mass renormalization  in field theories.

The Hamiltonian (\ref{Hamiltoniana}) is transformed to the principal
axis by means of a point transformation,
\begin{equation}
q_\mu = \sum_{r=0}^{N}t_\mu ^rQ_r \;\; , \, p_\mu
=\sum_{r=0}^{N}t_\mu ^rP_r\,, \label{transf}
\end{equation}
where $\mu =(0,\{k\})\, \;\;k=1,2,...,N$,
performed by an orthonormal matrix $T=(t_\mu ^r)$. The subscripts
$\mu =0$ and $\mu =k$ refer respectively to the atom and the
harmonic modes of the reservoir and $r$ refers to the normal
modes. In terms of normal momenta and coordinates, the transformed
Hamiltonian  reads
$$H=\frac 12\sum_{r=0}^N(P_r^2+\Omega _r^2Q_r^2),$$
where the $\Omega _r$'s are the normal frequencies corresponding
to the \textit{stable} collective oscillation modes of the coupled
system. It can be shown~\cite{adolfo1} that
\begin{equation}
t_k^r=\frac{c_k}{\omega _k^2-\Omega _r^2}t_0^r\;,\;\;t_0^r=\left[ 1
+ \sum_{k=1}^N\frac{c_k^2}{(\omega _k^2-\Omega _r^2)^2}\right]
^{-\frac 12}, \label{tkrg1}
\end{equation}
with the condition
\begin{equation}
\omega _0^2-\Omega _r^2 = \sum_{k=1}^N\frac{c_k^2}{\omega
_k^2-\Omega _r^2}. \label{Nelson1}
\end{equation}

To correctly describe the coupling of the atom with the field, we take
\begin{equation}
c_k=\eta \omega _k\,;\;\;\;\;\;\eta =\sqrt{2g\Delta \omega},
\label{eta}
\end{equation}
where $g$ is a constant with dimension of frequency. It measures the
strength of the coupling; $\Delta \omega=\pi c/R$ is the interval
between two neighboring frequencies of the reservoir and frequencies
of the field modes are given by~\cite{adolfo1},
\begin{equation}
\omega_k=k\Delta \omega=k\frac{\pi c}{R}
\label{omegak}
\end{equation}
The sum in Eq.~(\ref{Nelson1})  diverges for $N\rightarrow \infty $.
This makes the equation meaningless, unless  a renormalization
procedure, analogous to mass renormalization in field theories, is
implemented~\cite{Thirring}. Adding and subtracting a term
$\eta^2\Omega_r^2$ in the numerators of the right hand side in
Eq.~(\ref{Nelson1}) we have
\begin{equation}
\bar{\omega}^2-\Omega_r ^2 = \eta ^2\Omega _r^{2}\sum_{k=1}^\infty
\frac{1}{ \omega _k^2-\Omega_r ^2}, \label{Nelson3}
\end{equation}
where we define the \textit{renormalized} frequency
\begin{equation}
\bar{\omega}^2=\omega _0^2-\delta \omega ^2=\lim _{N \rightarrow
\infty }(\omega_{0}^2 - N\eta^2). \label{omegabarra}
\end{equation}
We find that the addition of a counterterm $-\delta \omega ^2q_0^2$
in the Hamiltonian Eq.~(\ref{Hamiltoniana}) compensates the
divergence of $\omega _0^2$ in such a way as to leave a finite,
physically meaningful renormalized frequency $\bar{\omega}$.

Using the formula,
\begin{equation}
\sum_{k=1}^{\infty}\frac{1}{(k^{2}-u^{2})} =
\left[\frac{1}{2u^{2}}-\frac{\pi}{u} {\rm cot}(\pi u)\right],
\label{id4}
\end{equation}
Eq.~(\ref{Nelson3}) can be rewritten as (dropping the label for the
eigenfrequencies),
\begin{equation}
\cot \left( \frac{R\Omega }c\right) =\frac{\Omega }{2\pi g}+\frac c{
2R\Omega }\left( 1-\frac{R\bar{\omega}^2}{\pi gc}\right) .
\label{cota}
\end{equation}
This gives an infinity of solutions. The  spectrum  of the
collective normal modes is denoted by $\Omega
_r\,;\;\;r=0,1,2,\cdots$. The transformation matrix elements
are~\cite{adolfo1},
\begin{eqnarray}
t_0^r & = &\frac{\eta \Omega _r}{\sqrt{(\Omega
_r^2-\bar{\omega}^2)^2+\frac{
\eta ^2}2(3\Omega _r^2-\bar{\omega}^2) + \pi ^2g^2\Omega _r^2}},
\nonumber \\
t_k^r &=&\frac{\eta \omega _k}{\omega _k^2-\Omega _r^2}t_0^r.
\label{t0r2}
\end{eqnarray}
Unless explicitly stated, the limit $N\rightarrow \infty $ is
understood in the following.

\section{Dressed states}

Let us consider the eigenstates of our system, $\left|
n_0,n_1,n_2...\right\rangle $, represented by the normalized
eigenfunctions in terms of the normal coordinates $\{Q_r\}$,
\begin{equation}
\phi _{n_0n_1n_2...}(Q)=\prod_s\left[ \sqrt{\frac{2^{n_s}}{n_s!}}
H_{n_s}\left( \sqrt{\frac{\Omega _s}\hbar }Q_s\right) \right]
~\Gamma _0,  \label{autofuncoes}
\end{equation}
where $H_{n_s}$ stands for the $n_s$-th Hermite polynomial and
$\Gamma _0$ is the normalized ground state eigenfunction,
\begin{equation}
\Gamma_{0}(Q) = {\cal{N}}\exp\left[-\frac{1}{2\hbar}
\sum_{r=0}^{\infty}\Omega_{r} Q_{r}^{2}\right]. \label{vacuo}
\end{equation}

We introduce \textit{dressed}  coordinates $q_0^{\prime }$ and
$\{q_k^{\prime }\}$ for, respectively, the \textit{dressed } atom
and the \textit{dressed} field, defined by,
\begin{equation}
\sqrt{\bar{\omega}_\mu }q_\mu ^{\prime }=\sum_rt_\mu
^r\sqrt{\Omega _r}Q_r, \label{qvestidas1}
\end{equation}
 where $\bar{\omega}_\mu
=\{\bar{\omega},\;\omega _k\}$. In terms of dressed coordinates, we
define for the time $\tau=0$, the \textit {dressed} states, $\left|
\kappa _0,\kappa _1,\kappa _2...\right\rangle $ by means of the
complete orthonormal functions
\begin{equation}
\psi _{\kappa _0\kappa _1...}(q^{\prime })=\prod_\mu \left[
\sqrt{\frac{ 2^{\kappa _\mu }}{\kappa _\mu !}}H_{\kappa _\mu }\left(
\sqrt{\frac{\bar{ \omega}_\mu }\hbar }q_\mu ^{\prime }\right)
\right] \Gamma _0, \label{ortovestidas1}
\end{equation}
where $q_\mu ^{\prime } = \left\{ q_0^{\prime },\,q_k^{\prime
}\right\} $, $ \bar{\omega}_\mu =\{\bar{\omega},\,\omega _k\}$.
Notice that the ground state, $ \Gamma _0$, in the above equation is
the same as in Eq.~(\ref{autofuncoes}). The invariance of the ground
state is due to our definition of dressed coordinates given by
Eq.~(\ref{qvestidas1}). In fact, we get the normal coordinates
$Q_{r}$ in terms of the dressed ones from Eq.~(\ref{qvestidas1}).
Replacing them in Eq.~(\ref{vacuo})  we find that the ground state
in terms of the {\it dressed} coordinates has the form
\begin{equation}
\Gamma_{0}(q') = {\cal{N}} \exp \left[ -\frac{1}{2\hbar}
\sum_{\mu=0}^{\infty} \bar{\omega}_\mu q_\mu'^2 \right] .
\label{ad2}
\end{equation}
Each function $\psi _{\kappa _0\kappa
_1...}(q^{\prime })$ describes a state in which the dressed
oscillator $q_\mu ^{\prime }$ is in its $\kappa _\mu $-th excited
state.

It is worthwhile to note that our dressed coordinates are new
objects, different from both the bare coordinates, $q$, and the
normal coordinates $Q$. In particular, the dressed  states, although
being collective objects, should not be confused with the
eigenstates given by Eq.~(\ref{autofuncoes}). While the eigenstates
$\phi$ are stable, all the dressed states $\psi$ are unstable,
except for the ground state $\Gamma_{0}$. The important idea is that
the dressed states are  physically meaningful.

In this framework, we write the physical states in terms of {\it
dressed} annihilation and creation operators $a_\mu'$ and
$a_\mu'^{\dag}$ defined in terms of dressed coordinates and momenta
in the usual way,
\begin{eqnarray}a_\mu'&=&\sqrt{\frac{\bar{\omega}_\mu}{2}}q_\mu'+
\frac{i}{\sqrt{2\bar{\omega}_\mu}}p_\mu'
\label{a1}\\
a_\mu'^\dag&=& \sqrt{\frac{\bar{\omega}_\mu}{2}}q_\mu'-
\frac{i}{\sqrt{2\bar{\omega}_\mu}}p_\mu'\;. \label{q4}
\end{eqnarray}
Then the initial dressed  density operator
corresponding to the thermal bath is given by
\begin{equation}
\rho'_\beta = \frac{1}{Z^{\prime}_\beta}
\exp\left[-\hbar\beta\sum_{k=1}^{\infty}\omega_k
\left(a_k'^{\dag}a_k'+\frac{1}{2} \right)\right], \label{ec6}
\end{equation}
with $Z^{\prime}_{\beta}=\prod_{k}z_{\beta}^{k\prime}$ being the
partition function of the dressed reservoir, where
\begin{equation} z_\beta^{k\prime}= {\rm
Tr}\left[e^{-\hbar\beta\omega_k\left(a_k'^\dag
a_k'+1/2\right)}\right]. \label{pfun}\end{equation} The system
evolves with time ($\tau$). The time-dependent {\it dressed}
occupation numbers are  defined as $$n_{\mu}'(\tau)={\rm
Tr}\left(a_\mu'^\dag(\tau)a_\mu'(\tau)\rho_0'\otimes\rho_\beta'\right)$$
(the prime is to clearly distinguish the dressed quantities from the
bare ones), where $\rho_0'$ is the density operator for the dressed
atom; $a_\mu'(\tau)$ and $a_\mu'^\dag(\tau)$ are the time-dependent
creation and annihilation operators.

The time evolution of the dressed annihilation operator is given by,
\begin{equation}
\frac{d}{d\tau}a_\mu'(\tau) =
i\left[{H},a_\mu'(\tau)\right]\; \label{v1}
\end{equation}
and a similar equation for $a_\mu'^\dag(\tau)$. We solve
this equation with the initial condition at time $\tau=0$,
\begin{equation}
a_\mu'(0)=\sqrt{\frac{\omega_\mu}{2}}q_\mu'+
\frac{i}{\sqrt{2\omega_\mu}}p_\mu'.
 \label{v2}
\end{equation}
which, in terms of bare coordinates, becomes
\begin{equation}
a_\mu'(0)=\sum_{r,\nu=0}^N\left(\sqrt{\frac{\Omega_r}{2}}
t_\mu^rt_\nu^{r}q_\nu
+\frac{it_\mu^rt_\nu^r}{\sqrt{2\Omega_r}}p_\nu\right)\;.
\label{q10}
\end{equation}
We assume a solution for $a_\mu'(\tau)$ of the type
\begin{equation}
a_\mu'(t)=\sum_{\nu=0}^{\infty}\left(\dot{B}_{\mu\nu}'(\tau){q}_\nu+
B_{\mu\nu}'(\tau){p}_\nu\right)\;. \label{v3}
\end{equation}
Using Eq.~(\ref{Hamiltoniana}) we find,
\begin{equation}
B_{\mu\nu}'(\tau) =
\sum_{r=0}^{\infty}t_\nu^r\left(a_{\mu}'^re^{i\Omega_r \tau}+
b_{\mu}'^re^{-i\Omega_r \tau}\right)\;. \label{B1}
\end{equation}
The initial conditions for $B_{\mu\nu}'(\tau)$ and
$\dot{B}_{\mu\nu}'(\tau)$ are obtained by setting $t=0$ in
Eq.~(\ref{v3}) and comparing with Eq.~(\ref{q10}); then
\begin{eqnarray}
B_{\mu\nu}'(0)& = &
i\sum_{r=0}^{\infty}\frac{t_\mu^rt_\nu^r}{\sqrt{2\Omega_r}}\;,
\label{q11}\\
\dot{B}_{\mu\nu}'(0)& = &
\sum_{r=0}^{\infty}\sqrt{\frac{\Omega_r}{2}}t_\mu^rt_\nu^r\;.
\label{q12} \end{eqnarray} Using these initial conditions and the
orthonormality of the matrix $\{t_\mu^r\}$ we obtain $a_\mu'^r=0$,
$b_\mu'^r=it_\mu^r/\sqrt{2\Omega_r}$. Replacing these values for
$a_\mu'^r$ and $b_\mu'^r$ in Eq.~(\ref{B1}) we get
\begin{equation}
B_{\mu\nu}'(\tau)=i\sum_{r=0}^{\infty}\frac{t_\mu^rt_\nu^r}
{\sqrt{2\Omega_r}}e^{-i\Omega_r t}\;. \label{q13}
\end{equation}
We have
\begin{eqnarray} a_\mu'(\tau)&=&\sum_{r,\nu=0}^{\infty}t_\mu^rt_\nu^r
\left(\sqrt{\frac{\Omega_r}{2}}{q}_\nu
+\frac{i}{\sqrt{2\Omega_r}}{p}_\nu\right)e^{-i\Omega_r \tau}\nonumber\\
& = &
\sum_{r,\nu=0}^Nt_\mu^rt_\nu^r\left(\sqrt{\frac{\omega_\nu}{2}}{q}_\nu'+
\frac{i}{\sqrt{2\omega_\nu}}{p}_\nu'\right)e^{-i\Omega_r \tau}\nonumber \\
&=&\sum_{\nu=0}^{\infty}f_{\mu\nu}(\tau){a}_\nu'\;,
\label{q14}
\end{eqnarray}
where
\begin{equation}
f_{\mu\nu}(\tau)=\sum_{r=0}^{\infty} t_\mu^rt_\nu^re^{-i\Omega_r
\tau}\, \label{fmunu}
\end{equation}
with $\mu\,,\nu=0,\,\{k\}$, $k=1,2,\cdots$.

This leads to the time evolution equation for the {\it dressed}
occupation number of the atom~\cite{termalizacao,livro},
\begin{equation}
n_{0}'(\tau) = |f_{0 0}(\tau)|^2n_0'+\sum_{k=1}^{\infty}|f_{0
k}(\tau)|^2n_k'\,, \label{q18}
\end{equation}
where  $n_0'$ stands for the occupation number at $\tau=0$.

\section{Thermal effects in a small cavity}

In this section we consider the {\it weak coupling} regime, defined
by
\begin{equation}
g=\bar{\omega}\alpha ,
\label{weak},
\end{equation}
where $\alpha $ is the fine-structure constant.

With $\eta =\sqrt{2g\pi c/R}$ and defining the
dimensionless parameter
\begin{equation}
\delta =\frac{g}{\Delta \omega} =\frac{gR}{\pi c},
\label{delta}
\end{equation} Eq.~(\ref{cota}) becomes,
\begin{equation}
\cot \left(\frac{\pi \Omega \delta }{g}\right) = \frac{\Omega }{2\pi
g}+\frac{ g}{2\pi \delta \Omega }\left( 1-\frac{\delta
\bar{\omega}^2}{g^2}\right) . \label{cota1}
\end{equation}

Let us consider  the right hand side of Eq.~(\ref{cota1})  such that
\begin{equation}
\frac{\delta \bar{\omega}^2}{g^2}>1.
\label{condicao1}
\end{equation}
In the weak coupling regime,  this corresponds to a value of
$\delta$, $\delta\gtrsim 5.3\times 10^{-5}$. For a frequency
$\bar{\omega}=4.0\times 10^{14}/s$ (in the visible red) this gives a
condition on the cavity size of $R\gtrsim 1.7\times 10^{-8}\,m$.
Then the general behavior of the solutions of Eq.~(\ref{cota1}) is
illustrated in Fig(\ref{cotangente}). We find that all but one of
the eigenfrequencies are very close to the frequencies of the field
modes, $\omega_k$, given by Eq.~(\ref{omegak}). Then we label
solutions for the eigenfrequencies $\Omega_r$ as, $\Omega_0$,
$\Omega_k$, $k=1,2,\cdots$.  The solutions $\Omega_k$ of
Eq.~(\ref{cota1}) are,
\begin{equation}
\Omega _k = \frac{\pi c}R\left( k+\epsilon _k\right)\,\;\;k=1,2,\ldots \,,
\label{OmegaK}
\end{equation}
with $0<\epsilon_{k}<1$, satisfying the equation,
\begin{equation}
\mathrm{cot}(\pi \epsilon_k) = \frac{2c}{gR}(k+\epsilon_k) +
\frac{1}{(k+\epsilon_k)} \left(1-\frac{\bar{\omega}^{2}R}{\pi
gc}\right). \label{eigen2}
\end{equation}
Since every $\epsilon_k$ is  much smaller than $1$,
Eq.~({\ref{eigen2}) can be linearized in $\epsilon_k$, giving,
\begin{equation}
\epsilon_k=\frac{\pi g c R k}{\pi^{2} c^{2} k^{2}-\bar{\omega}^{2}R^{2}}.
\label{linear}
\end{equation}
The eigenfrequencies, $\Omega_{k}$, are obtained by solving
Eqs.~(\ref{OmegaK}) and (\ref{eigen2}) or (\ref{linear}).

\begin{figure}[t]
\includegraphics[{height=6.0cm,width=8cm,angle=360}]{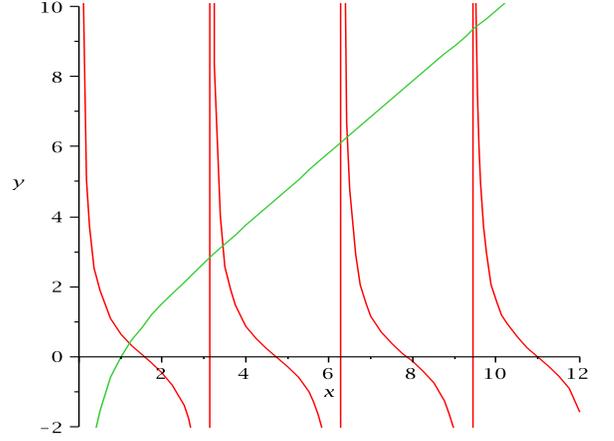}
\caption{Solutions of Eq.~(\ref{cota1}), for cavities satisfying the
condition Eq.~(\ref{condicao1}); the asymptotes of the cotangent
curve correspond to the frequencies of the field modes $\omega_k$.}
\label{cotangente}
\end{figure}

The lowest eigenfrequency, $\Omega_{0}$, is obtained by  assuming
that it satisfies the condition $\Omega_{0}R/c\ll 1$. Inserting this
condition in Eq.~(\ref{cota1}) and keeping up to quadratic terms in
$\Omega$ we obtain the solution for the lowest eigenfrequency,
\begin{equation}
\Omega _0\approx \bar{\omega}\left( 1-\frac{\pi \delta }{2}\right) .
\label{Omega0}
\end{equation}
Consistency between Eq.~(\ref{cota1}) and the condition
$\Omega_{0}R/c\ll 1$ gives a condition on $R$, $i.e$ $R\ll (c/g)
\lambda$, with $\lambda = (\pi/2)\left(g/\bar{\omega}\right)^2$.

Let us first determine the temperature independent term
$|f_{00}(\tau)|^2n_0'$ in Eq.~(\ref{q18}), considering that the dressed atom is initially (at $\tau=0$) in the first excited level, that is $n_0'=1$. We evaluate  $(t_0^0)^2$ and
$(t_0^k)^2$ from Eqs.~(\ref{t0r2}), (\ref{OmegaK}), (\ref{linear}),
and (\ref{Omega0}) to find
\begin{equation}
(t_0^k)^2 \approx \frac{2gR}{\pi c k^2} = \frac{2 \delta}{k^2} ,
\label{too}
\end{equation}
and then using the normalization condition
$\sum_{r=0}^{\infty}(t_0^r)^2=1$ and $\zeta(2) = \sum_{k=1}^{\infty}
k^{-2} = \pi^2 /6$, we have
\begin{equation}
(t_0^0)^2 \approx 1-\frac{\pi g R}{3c} = 1 - \frac{\pi^2 \delta}{3}
\label{too1}
\end{equation}
From Eq.~(\ref{fmunu}), using the de Moivre formula,
$e^{i\theta}=\cos\theta+i\sin\theta$, we have
\begin{equation}
|f_{\mu\nu}(\tau)|^{2} = \sum_{r,s=0}^{\infty} t_\mu^r t_\nu^r
t_\mu^s t_\nu^s \cos(\Omega_r -\Omega_s )\tau. \label{demoivre}
\end{equation}
Let us assume that the thermal bath is at zero temperature, $i.e.$ all the modes of the reservoir are in the ground state, $n_k'=0$ for all values of $k$. Taking the above approximations for $t_0^k$ and $t_0^0$, we get from Eq.~(\ref{q18}), the zero-temperature  time evolution of the occupation number of the dressed atom initially in the first excited level, 
\begin{widetext}
\begin{equation}
|f_{00}(\tau)|^{2} \approx \left(1-\frac{\pi^2
\delta}{3}\right)^2+4\delta\left(1-\frac{\pi^2 \delta}{3}\right)
\sum_{k=1}^{\infty}\frac{1}{k^{2}}\cos(\Omega_{k}-\Omega_{0})\tau+
4\delta^{2}\sum_{k,l=1}^{\infty} \frac{1}{k^{2}l^{2}}\cos
(\Omega_{k}-\Omega_{l})\tau. \label{rho11}
\end{equation}
\end{widetext}
This is an oscillating function which has a minimum value
$\mathrm{Min}(|f^{00}(t)|^{2})$. Taking both cosine functions in
Eq.~(\ref{rho11}) equal to $-1$, we get a lower bound for
$\mathrm{Min}(|f^{00}(t)|^{2})$ given, up to first order in
$\delta$, by
\begin{equation}
F(\delta)=1-\left(\frac{2\pi^2}{3}-2\right)\delta.
\label{rho11ss}
\end{equation}
As an example we consider that the atom in the first excited state has an emission frequency  $\bar{\omega}\approx 4\times 10^{14}/s$,
in the visible red, and we take the radius of the confining cavity $R\approx 10^{-6}$m. With these data we get
$F(\delta)\approx 0.99$, that is a probability of $99\%$ at zero
temperature, that it will almost never decay. This shows the high
stability of the system, which is confirmed by
experiment~\cite{Haroche3,Hulet}.

In order to take into account the temperature effects, we must
consider the second term in Eq.~(\ref{q18}), that is we must
evaluate the quantity
\begin{equation}
|f_{0 k}(\tau)|^2 = |t_0^0t_k^0 e^{-i\Omega_{0}\tau} +
\sum_{l=1}^{\infty}t_0^lt_k^le^{-i\Omega_{l}\tau}|^2.
\label{f0kpequeno}
\end{equation}
This is carried out by using the matrix elements obtained from
Eq.~(\ref{t0r2}) and the formulas for eigenfrequencies in a small
cavity.

It is assumed that the thermal distribution of the occupation
numbers of the field modes in the cavity  follow the Bose-Einstein
distribution,
\begin{equation}
n_k^{\prime}=\frac{1}{e^{\hbar\beta\omega_k}-1}.
\label{nk}
\end{equation}
This can be justified in the following way: in the case of an
arbitrarily large cavity, the dressed field modes coincide with the
bare ones~\cite{adolfo1}, which in the limit of vanishing coupling
makes this approximation exact. Strictly speaking this is not the
case for a finite cavity. Nevertheless, in many situations this
approximation is acceptable in the weak coupling r\'egime. For
instance a cavity of radius $R\approx 10^{-6}m$ is  $\sim 10^{4}$
times larger than the size of a hydrogen atom (the Bohr radius). In
such a case the atom ``sees" the cavity to be a very large one and
the approximation is justified. Then from Eqs.~(\ref{f0kpequeno})and
(\ref{q18}), we get the time evolution of the temperature dependent
occupation number for the atom,
\begin{widetext}
\begin{equation}
n_{0}'(\tau,\beta) = |f_{0 0}(\tau)|^2n_0' +
\sum_{k=1}^{\infty}\frac{1}{e^{(\hbar\beta \pi
c/R)k}-1}\left[(t_0^0)^2(t_k^0)^2 +
2\sum_{l=1}^{\infty}t_0^0t_0^lt_k^0t_k^l\cos(\Omega_0-\Omega_l)\tau
+ \sum_{l,n=1}^{\infty} t_0^lt_0^nt_k^lt_k^n
\cos(\Omega_l-\Omega_n)\tau \right] \nonumber \\
\label{n0tbeta}
\end{equation}
\end{widetext}
where $|f_{0 0}(\tau)|^2$ is given by Eq.~(\ref{rho11}). The matrix
elements $t_{k}^{0}$ and $t_{k}^{l}$ in the above formulas are
evaluated from Eqs.~(\ref{t0r2}), (\ref{OmegaK}) and (\ref{Omega0})
to be,
\begin{equation}
t_{k}^{0} = \frac{k\,g^2\sqrt{2\delta}}{k^2g^2-\Omega_0^2\delta^2}\,;\;\;
t_{k}^{l} = \frac{2k\delta}{k^2-(l+\epsilon_{l})^2}\frac{1}{l} .
\nonumber \\
\label{ts}
\end{equation}
The occupation number $n_{0}'(\tau,\beta)$ is an oscillating
function which has a minimum value,
${\rm{Min}}[n_{0}'(\tau,\beta)]$, that depends on the temperature
$\beta^{-1}$. We can obtain a lower bound, $n_{0}'(\beta)$, for this
minimum, such that ${\rm{Min}}[n_{0}'(\tau,\beta)] > n_{0}'(\beta)$,
by taking both cosine functions in Eq.~(\ref{n0tbeta}) equal to $-1$
\begin{widetext}
\begin{equation}
n_{0}'(\beta) = F(\delta)n_0' +
\sum_{k=1}^{\infty}\frac{1}{e^{(\hbar\beta \pi c/R)k}-1}
 \left[(t_0^0)^2(t_k^0)^2
- 2\sum_{l=1}^{\infty}t_0^0t_0^lt_k^0t_k^l
-\sum_{l,n=1}^{\infty}t_0^lt_0^nt_k^lt_k^n \right].
\label{minn0}
\end{equation}
\end{widetext}
Numerical calculation of Eqs.~(\ref{n0tbeta}) and (\ref{minn0}) will
describe how the time evolution of the occupation number and the
stability of the excited atom are affected by heating. We take for
the plots $n_0'=1$, that is the atom initially in the first excited
level.
\begin{figure}[t]
\includegraphics[{height=6.0cm,width=8cm,angle=360}]{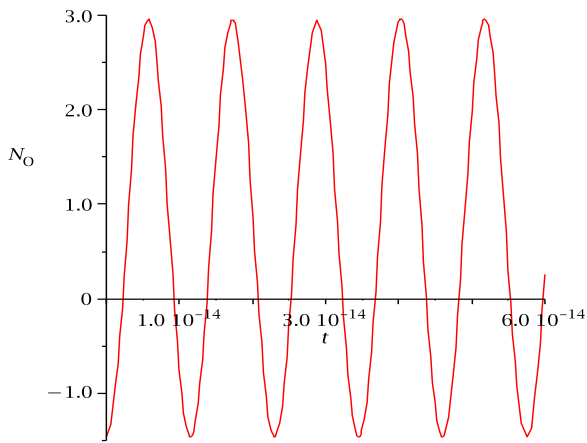}
\caption{Time evolution of the thermal dependent occupation number
$n_{0}'(\tau,\beta)$ for a temperature $T=300\,K\,$; the scale for
the vertical axis is $N_0=(n_{0}-1)\times 10^{-12}$; time is in
seconds.} \label{T300}
\end{figure}

\begin{figure}[t]
\includegraphics[{height=6.0cm,width=8cm,angle=360}]{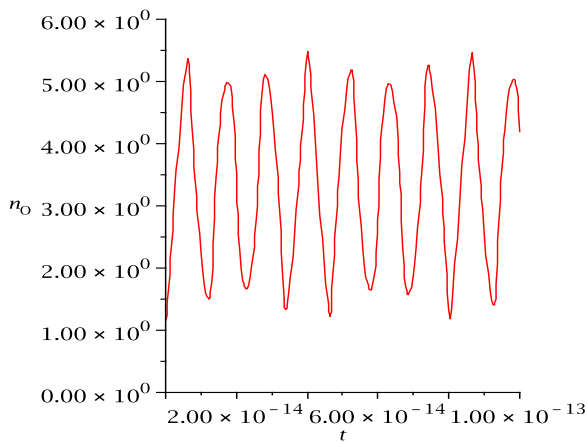}
\caption{Time evolution of the thermal dependent occupation number
$n_{0}'(\tau,\beta)$ for a temperature $T=10^{5}\,K$; time is in
seconds.} \label{T100000}
\end{figure}

\begin{figure}[t]
\includegraphics[{height=6.0cm,width=8cm,angle=360}]{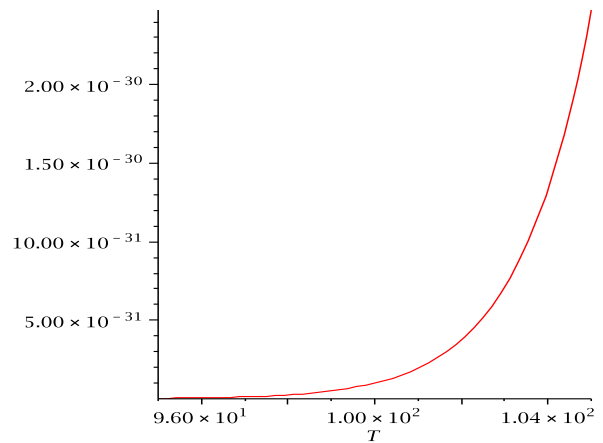}
\caption{Temperature behavior of the lower bound for the minimum of
$n_{0}'(\tau,\beta)$, $n_{0}'(\beta)$, given by Eq.~(\ref{minn0});
temperature is in Kelvin.} \label{T1min}
\end{figure}

In Fig.~(\ref{T300}) and in Fig.~(\ref{T100000}) the time evolution
of the temperature  dependent occupation number $n_{0}'(\tau,\beta)$
is plotted for some representative values of the emission frequency
and the temperature. In Fig.~(\ref{T1min}) the lower bound for its
minimum, $n_{0}'(\beta)$, is plotted as a function of temperature.
We find from these figures that raising temperature increases the
amplitude of oscillation of the occupation number and that its
minimum lower bound, $n_{0}'(\beta)$, also grows with temperature. For
a  given emission frequency $\bar{\omega}=4.0\times 10^{14}/s$, the
increase  of the amplitude of oscillation  of $n_{0}'(\tau,\beta)$
and of its lower bound $n_{0}'(\beta)$ are negligible for room temperatures, they
are significant for high laboratory temperatures. In
Fig.~(\ref{T100000}) $n_{0}'(\tau,\beta)$ is plotted for
$T=10^{5}\,K$ ($\sim 8.4\,eV$); although this temperature is very
high for everyday life, it can, in
principle, be attained in laboratory for excited atoms. In fact it is lower than the ionization temperature of $13.6\,eV$ for the hydrogen atom, and still much lower than the nuclear
fusion temperature of $\approx 10^{8}\,K$ ($\sim 8.4\,KeV$). We find that the average
occupation number at this temperature ($T=10^{5}\,K$) is about $3.5$ times higher
than the zero temperature value $n_{0}'(\tau,T=0)\approx 1$. At room
temperature the  occupation number will remain very close to the
zero-temperature value, as is shown in Fig.~(\ref{T300}). Therefore
we find that as the temperature is raised, both the amplitude of
oscillation of the occupation number and its minimum, grow with
respect to the zero-temperature values.

\section{Concluding remarks}

At zero temperature the dressed atom, initially in the first or
higher excited state, can only decay, since all field modes are in
the ground state. It is inhibited from decaying by confinement in a
cavity of small size. However at finite temperature, the field modes
in the cavity can be in excited states with a finite probability
given by the Bose-Einstein distribution function. As a consequence the dressed atom can exchange quanta from the field. This means that the thermal occupation
number of excited states of the atom increases with temperature. In other words, as an effect of heating  the atom will be in 
  a higher
excited state which may be able to decay. However the decay is inhibited by the confining geometry. The
results presented above give sufficient proof of these ideas.

This behavior is also to be contrasted with the situation of an
arbitrarily large cavity (free space) described
in~\cite{termalizacao,livro}. In that case, for long times the
dressed occupation number of the atom approaches smoothly to an
asymptotic value which is nearly the one  obtained from the Bose
distribution at the  equilibrium temperature of the reservoir.
Taking the same value as before for $\bar{\omega}$, this value is
$n_{0}'(t\rightarrow \infty,\beta;\bar{\omega},R\rightarrow
\infty)=1/(e^{14}-1)\approx 0$. In that case the growth of the
Bose-Einstein weight factor due to raising temperature is
compensated by the lowering due to larger and larger values of $R$,
leading to an equilibrium occupation number.

\section*{Acknowledgements}

A.P.C.M. and A.E.S. are grateful to the Theoretical Physics
Institute, University of Alberta, for kind hospitality during the
summer 2009. The research of F.C.K. is funded by NSERC (Canada).
A.P.C.M., J.M.C.M. and A.E.S. are supported by CNPq and CAPES
(Brazil).


\begin{thebibliography}{99}

\bibitem{farina1} C. Farina, T. N. C. Mendes, F. S. S. Rosa and A.
Tenorio, Phys. Rev. A {\bf 78}, 012105 (2008).

\bibitem{farina2} C. Farina and T. N. C. Mendes, J. Phys A: Math. Gen.
{\bf 40}, 7343 (2007).

\bibitem{bouquinCohen} C. Cohen-Tannoudji, "Atoms in Electromagnetic Fields",
(World Scientific, Singapure, 1994).

\bibitem{Wylie} J. M. Wylie and J. E. Sipe, Phys. Rev. A {\bf 30}, 1185
(1984).

\bibitem{Jhe} W. Jhe and K. Jang, Phys. Rev. A {\bf 53}, 1126
(1996).

\bibitem{adolfo1} N. P. Andion, A. P. C. Malbouisson and A. Mattos Neto, J. Phys.
A: Math. Gen. {\bf 34}, 3735, (2001).

\bibitem{adolfo2} G. Flores-Hidalgo, A. P. C. Malbouisson and Y. W. Milla,
Phys. Rev. A {\bf 65}, 063414 (2002).

\bibitem{adolfo3} G. Flores-Hidalgo and A. P. C. Malbouisson, Phys. Rev. A {\bf 66},
042118 (2002).

\bibitem{adolfo4}  A. P. C. Malbouisson, Annals of Physics \textbf{308}, 373
(2003).

\bibitem{adolfo5}  G. Flores-Hidalgo and A. P. C. Malbouisson, Phys. Lett. A
\textbf{337}, 37 (2005).

\bibitem{adolfo6} G. Flores-Hidalgo, C.A. Linhares, A.P.C.
Malbouisson and J.M.C. Malbouisson, J. Phys. A: Math. Ther.
\textbf{41}, 075404 (2008).

\bibitem{zurek}  W. G. Unruh and W. H. Zurek, Phys. Rev. D \textbf{40}, 1071
(1989).

\bibitem{paz}  B. L. Hu, J. P. Paz and Y. Zhang, Phys. Rev. D \textbf{45},
2843 (1992).

\bibitem{caldeira}  A. O. Caldeira and A. J. Leggett, Ann. Phys. (N.Y) \textbf{
149}, 374 (1983).

\bibitem{caldeira1} M. Rosenau da Costa, A. O. Caldeira, S. M. Dutra and H.
Westfahl Jr., Phys. Rev A {\bf 61}, 022107 (2000).

\bibitem{Thirring}  W. Thirring and F. Schwabl, Ergeb. Exakt. Naturw. {\bf 36},
219 (1964).

\bibitem{Haroche3} W. Jhe, A. Anderson, E. A. Hinds, D. Meschede, L. Moi and S.
Haroche, Phys. Rev. Lett. {\bf 58}, 666 (1987).

\bibitem{Hulet} R.G. Hulet, E. S. Hilfer and D. Kleppner, Phys. Rev. Lett.
{\bf 55}, 2137 (1985).

\bibitem{termalizacao} G. Flores-Hidalgo, A. P. C. Malbouisson, J. M. C.
Malbouisson, Y. W. Milla and A. E. Santana, Phys. Rev. A (Online),
{\bf 79}, 032105 (2009).

\bibitem{livro} F. C. Khanna, A. P. C. Malbouisson, J. M. C. Malbouisson and A. E.
Santana, {\it Thermal Quantum Field Theory - Algebraic Aspects and
Applications}, 1. ed. (World Scientific, Singapore, 2009).



\end{thebibliography}
\end{document}